\documentclass[journal=jctcce, manuscript=article]{achemso}

\usepackage[version=4]{mhchem}
\usepackage{float}
\usepackage{tabularx}
\usepackage{graphicx}
\usepackage{hyperref}
\usepackage{soul}
\usepackage{bm}
\usepackage{multirow}
\usepackage{makecell}
\usepackage{placeins}
\usepackage{amsmath}
\usepackage{flexisym}
\usepackage[table]{xcolor}
\usepackage{expl3}
\usepackage{booktabs}
\usepackage{comment}
\usepackage{physics}
\usepackage{amssymb,mathtools}
\usepackage{cleveref}

\author{Shreya Verma}
\author{Abhishek Mitra}
\affiliation[University of Chicago]
{Department of Chemistry, University of Chicago, Chicago, IL 60637, USA.}
\author{Qiaohong Wang}
\affiliation[University of Chicago]
{Pritzker School of Molecular Engineering, University of Chicago, Chicago, IL 60637, USA.}
\author{Ruhee D'Cunha}
\author{Bhavnesh Jangid}
\author{Matthew R. Hennefarth}
\author{Valay Agarawal}
\author{Leon Otis}
\author{Soumi Haldar}
\affiliation[University of Chicago]
{Department of Chemistry, University of Chicago, Chicago, IL 60637, USA.}

\author{Matthew R. Hermes}
\affiliation[University of Chicago]
{Department of Chemistry, University of Chicago, Chicago, IL 60637, USA.}
\author{Laura Gagliardi}
\email{lgagliardi@uchicago.edu}
\affiliation[University of Chicago]
{Department of Chemistry, University of Chicago, Chicago, IL 60637, USA.}
\alsoaffiliation[University of Chicago]
{Pritzker School of Molecular Engineering, University of Chicago, Chicago, IL 60637, USA.}
\alsoaffiliation[University of Chicago]
{Chicago Center for Theoretical Chemistry, University of Chicago, Chicago, IL 60637, USA.}

\title{Multireference Embedding and Fragmentation Methods for Classical and Quantum Computers: from Model Systems to Realistic Applications}
 
\begin{document}
\begin{tocentry}
\centering
\includegraphics[width=\textwidth]{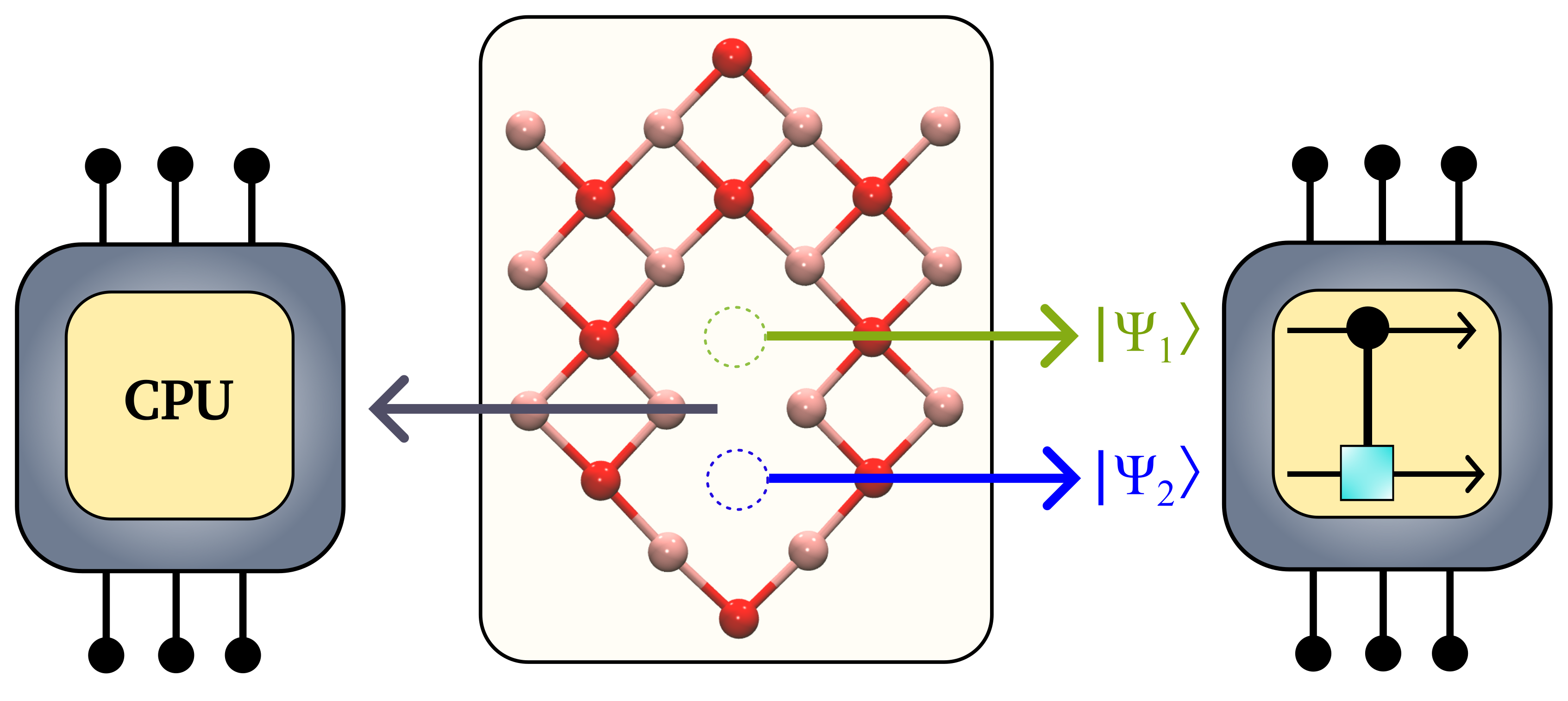}
\end{tocentry}

\begin{abstract}
	One of the primary challenges in quantum chemistry is the accurate modeling of strong electron correlation. 
    While multireference methods effectively capture such correlation, their steep scaling with system size prohibits their application to large molecules and extended materials.
    Quantum embedding offers a promising solution by partitioning complex systems into manageable subsystems. 
    In this Account, we highlight recent advances in multireference density matrix embedding and localized active space self-consistent field approaches for complex molecules and extended materials. 
    We discuss both classical implementations and the emerging potential of these methods on quantum computers. 
    By extending classical embedding concepts to the quantum landscape, these algorithms have the potential to expand the reach of multireference methods in quantum chemistry and materials.
\end{abstract}

\section{Introduction}

Quantum simulations are essential for addressing challenges across various fields, including catalysis\cite{bell2011,krylov2018,xu2019}, quantum information\cite{dreyer2018,wasielewski2020,wolfowicz2021quantum}, and materials science~\cite{maurer2019advances}. They enable the modeling of reaction mechanisms, the optimization of catalysts, and the exploration of quantum system behavior. Achieving practical relevance requires simulating large systems, which is a challenge due to the significant computational cost of quantum mechanical methods.

Many of these systems exhibit locality, where key physical or chemical phenomena are confined to a specific region. Quantum embedding methods leverage this locality by partitioning a system into smaller, high-accuracy subsystems and larger, low-cost environments. Quantum embedding approaches are usually classified based on the quantum variable of interest \cite{sun2016quantum,jones2020,vorwerk2022}, such as the density \cite{cortona1991,wesolowski1993,elliott2009,elliott2010,jacob2014,libisch2014,wesolowski2015,wesolowski2008embedding,pernal2009orbital,wesolowski2020correlation,wesolowski2004hydrogen,wesolowski2014embedding}, Green’s function \cite{kotliar2006,zgid2011,chibani2016,zgid2017,lan2017,ma2021,sheng2022,chen2025advances,rusakov2019}, or density matrix~\cite{knizia2012density,knizia2013density,wouters2016,5years_of_DMET,negre2025new,sekaran2021householder,sekaran2023unified,marecat2023unitary}. 

When the quantum variable of interest is the density, such as in density functional theory (DFT)-in-DFT\cite{cortona1991,wesolowski1993,wesolowski2015,wesolowski2008embedding,pernal2009orbital,wesolowski2020correlation,wesolowski2004hydrogen,wesolowski2014embedding} and wave function (WF)-in-DFT embedding \cite{govind1998,govind1999,huang2011,manby2012,goodpaster2012,chulhai2018,lee2019,kolodzeiski2023automated}, the density is partitioned in real space between the region of interest and the environment.
An embedding potential is defined either through a functional derivative of the energy with respect to the subsystem density \cite{govind1999} or by maximizing the Wu-Yang-like functional with respect to the embedding potential itself~\cite{huang2011}.
Often, the potential and corresponding density can be obtained self-consistently with the requirement that the subsystem densities sum to the total density~\cite{huang2006self}.
For DFT-in-DFT embedding, computational savings are obtained through the use of
different levels of DFT, particularly those with and without exact exchange, for different parts of the system~\cite{fornace2015}.
On the other hand, in the case of WF-in-DFT, the region of interest is treated using a correlated WF and the surrounding environment using DFT, with a local embedding potential to connect the different methods in different regions~\cite{govind1998,govind1999,huang2011,goodpaster2012}. The potential field embedding theory involves a fully self-consistent hybrid correlated WF/DFT calculation that enables the interactions between different subsystems to go beyond the DFT level~\cite{huang2011,cheng2017potential1,cheng2017potential2}. Determining the conditions for updating the embedding potential self-consistently in embedding approaches remains an open question. One of the central challenges of density-based embedding approaches is the accurate removal of double counting errors wherein some correlation energy of the fragment is included from both the lower level DFT and the WF treatment~\cite{kotliar2006,karolak2010double,haule2015exact,sheng2022}.

Embedding methods based on the Green’s function have a long history of development for both model Hamiltonians\cite{georges1992} and \textit{ab initio} calculations~\cite{zgid2011,lin2011,turkowski2012,lan2015}.
These techniques account for electronic interactions through the self-energy and employ a partitioning of contributions to the self-energy between different levels of theory.
Notable methods within this category include dynamical mean field theory (DMFT) \cite{anisimov1997,lichtenstein1998,biermann2003}, self-energy embedding theory (SEET)\cite{lan2015,lan2017,zgid2017,rusakov2019}, and quantum defect embedding theory (QDET)~\cite{ma2021,sheng2022,chen2025advances}.
Broadly, these methods differ in the choices of low-level and high-level theories, active space selection schemes, and double counting corrections. The use of either DFT or GW (Green function G and a
screened Coulomb interaction W) as the low-level method has been a common choice within the DMFT framework (often denoted DFT+DMFT\cite{lichtenstein1998} or GW+DMFT\cite{biermann2003,biermann2014}) and recent work with QDET has formulated an exact double counting correction at the G0W0 level~\cite{sheng2022}. Applications of SEET\cite{kananenka2015,lan2015,lan2017,rusakov2019} generally employ the second order Green’s function method (GF2) as the low-level theory and most of the methods commonly use full configuration interaction (FCI) as the high-level theory. Active-space orbitals may be selected by hand (e.g. 3d orbitals in many applications of DMFT)\cite{kotliar2006,nilsson2018}, through diagonalization of the 1-body density matrix as in SEET \cite{lan2015,lan2017}, or according to criteria like spatial localization factor \cite{sheng2022} or Kohn-Sham energy in QDET~\cite{chen2025advances}.
There are also methods that embed electronic structure theory into random phase approximation (RPA) \cite{schafer2021local,schafer2021surface}, RPA embedding in DFT~\cite{wei2023introducing}. Sometimes downfolding techniques are employed in which the weakly interacting environment is downfolded on the correlated subspace with efficient stochastic constrained RPA~\cite{romanova2023dynamical}. 

Auxiliary field quantum Monte Carlo \cite{lee2022twenty} also offers embedding strategies where a local region of the system is embedded in an environment frozen at the independent-electron level~\cite{virgus2014stability,motta2018ab}. Active space embedding theory is another approach for embedding a multireference method within a environment treated at a frozen-orbital level~\cite{he2020zeroth,he2022second}.

\begin{figure}[H]
	\centering
	\includegraphics[width=\textwidth]{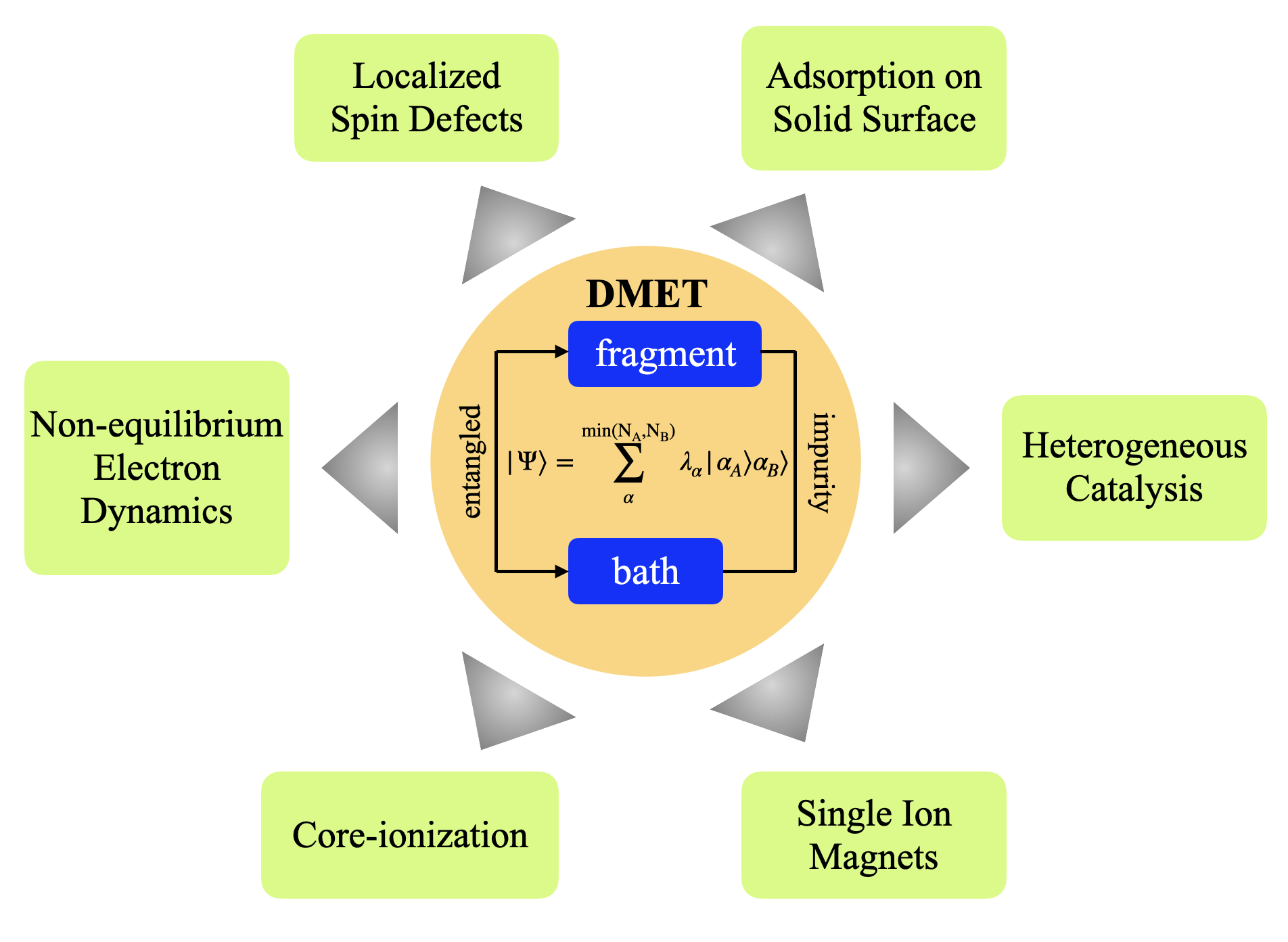}
	\caption{Overview of potential applications of  density matrix embedding theory across various domains in quantum chemistry.}
	\label{fig:DMET-cat-1}
\end{figure}

Density matrix embedding theory (DMET) was historically motivated as a conceptually simpler and computationally efficient alternative to DMFT~\cite{knizia2012density}. It was soon extended from model Hamiltonians to fully \textit{ab initio} problems~\cite{knizia2013density}. \Cref{fig:DMET-cat-1} highlights key applications for DMET, which usually involve close lying electronic states. These include point defects in solid state systems \cite{bockstedte2018ab}, spin-state energetics in transition metal complexes \cite{harvey2003understanding}, magnetic molecules \cite{pederson2019multireference}, and molecule-surface interactions \cite{mosquera2020quantum}; all of which are characterized by strong electron correlation. To address such challenges, DMET has recently been integrated with active-space multireference quantum eigensolvers, such as the complete active space self-consistent field (CASSCF) method~\cite{RoosComplete1980, RoosComplete1987}.

Existing active-space methods, whether used with or without an embedding scheme, have an exponential scaling, making calculations with active spaces beyond 20 electrons and 20 orbitals prohibitive \cite{Vogiatzis2017}. Quantum computing provides a promising alternative\cite{mcardle_quantum_2020} to classical computers due to its theoretical polynomial scaling in the simulation of inherently quantum mechanical systems \cite{QAforQC_rev, QCinQC_rev, Poly} and potential for exponential speedups with algorithms like quantum phase estimation (QPE)~\cite{QPE,QPE_ExpoSpeedUp,lee_evaluating_2023}. Despite this potential, the realization of ``quantum advantage" in these specific domains remains largely unachieved. This is primarily due to limitations in current quantum computing hardware, such as noisy qubits, restricted qubit entanglement, decoherence errors, readout inaccuracies, and errors in quantum gate implementation~\cite{Preskill2018}. Hybrid approaches such as the variational quantum eigensolver (VQE)\cite{VQE} have shown success in leveraging quantum solvers for systems with small numbers of electrons and orbitals. In the noisy intermediate-scale quantum computing (NISQ) era, quantum embedding theories are useful to reduce the complexity of electronic structure calculations of extended systems. Integrating quantum embedding methods \cite{Ma2020,vorwerk2022,galli,bootstrap,Rossmannek2023,Cao2023} with quantum solvers, such as using VQE for the subsystem and classical methods for the environment, reduces complexity and can help push the boundaries of what is currently possible with classical multireference algorithms.

This Account focuses on density matrix embedding theory, its recent multireference extensions, and its integration with quantum algorithms, particularly in the context of localized active space methods, and finally offers an outlook on the future development of the field.

\section{Density matrix embedding theory and recent advancements}
\subsection{Density matrix embedding theory (DMET)}
The non-relativistic electronic molecular Hamiltonian in second quantization is:
\begin{equation}
    \label{eq:mol-ham}
	\hat{H}=\sum_{pq}^{N}h_{pq} \hat{E}_{pq} + \frac{1}{2}\sum_{pqrs}^{N}V_{pqrs}\hat{e}_{pqrs} + V_{NN}
\end{equation}

Here, $\hat{E}_{pq}$ and $\hat{e}_{pqrs}$ are the spin-summed one- and two-electron excitation operators, $h_{pq}$ and $V_{pqrs}$ are the one- and two-electron integrals in a spatial orbital basis, and $V_{NN}$ is the nuclear repulsion energy. Accurately solving this Hamiltonian using standard electronic structure methods scales either polynomially $\mathcal{O}(N^x)$ or exponentially $\mathcal{O}(e^N)$ with $N$. DMET reduces this scaling by partitioning a system into one or more finite fragments and their environment, allowing entanglement between the sub-parts. 

The DMET algorithm begins with a converged mean-field wave function, primarily the Hartree-Fock (HF) wave function, denoted as $\ket{\phi_{\mathrm{HF}}}$, rather than DFT to prevent double counting errors~\cite{wouters2016}. The orbitals are subsequently localized on atomic centers through methods such as Pipek-Mezey (PM) \cite{pipek1989fast}, Boys \cite{foster1960canonical}, and meta-L\"owdin~\cite{lowdin1950non}. Using the localized orbitals, the wave function of the entire system is partitioned into a fragment $A$ (a user-selected subset of atoms) and its environment $B$ (with $N_A$ and $N_B$ representing the number of orbitals in the subspaces A and B, respectively). This partitioning allows the wave function in the combined space of A and B to be expressed in a product form:
\begin{equation}
	\ket{\Psi} = \sum_i^{N_A}\sum_j^{N_B}C_{ij}\ket{i_A}\ket{j_B}
\end{equation}
with $\ket{i_A}$, $\ket{i_B}$ being the localized states centered on the fragments $A$, $B$, respectively. The singular value decomposition of the coefficient tensor $C_{ij}$ yields basis transformation matrices $U_{i\alpha}$ and $V^\dagger_{\alpha j}$:
\begin{equation}
C_{ij}=\sum_\alpha^{\min(N_A,N_B)}U_{i\alpha}\lambda_\alpha V^\dagger_{\alpha j} \implies |\Psi\rangle = \sum_\alpha^{\min(N_A,N_B)}\lambda_\alpha|\alpha_A\rangle|\alpha_B\rangle
\label{eq:svd_eq}
\end{equation}

The wave function is transformed to the rotated basis $\ket{\alpha_A}\ket{\alpha_B}$, where the singular values $\lambda_\alpha$ indicate the degree of entanglement. The new states $\ket{\alpha_B}$ represent the DMET bath states, capturing entanglement with the environment. The embedding Hamiltonian is then constructed by projecting the full Hamiltonian, $\hat{H}$, onto the fragment and bath states $\ket{\alpha_A}, \ket{\alpha_B}$:
\begin{equation}
	\hat{H}_{\mathrm{emb}}=\sum_{pq}^{2N_A}\tilde{h}_{pq}\hat{E}_{pq} + \frac{1}{2}\sum_{pqrs}^{2N_A}\tilde{V}_{pqrs}\hat{e}_{pqrs}
\end{equation}

where $\tilde{h}_{pq}$ and $\tilde{V}_{pqrs}$ are the embedding-states projected one- and two-particle integrals. Note that the sum now runs over $2N_A$ because there can be a maximum of $N_A$ bath states that restricts the embedding Hamiltonian to an embedding space of dimension $2N_A$. The embedding Hamiltonian can then be solved with high-level quantum chemical methods based on the problem's requirements. In general, the system can be partitioned into any number of non-overlapping fragments. For more than a single fragment, the Hamiltonian of the full system needs to be augmented with a global chemical potential $\mu_{\mathrm{glob}}$ to ensure the sum of electrons within the fragments adds up to the total electrons in the system. A correlation potential ($u$) for each fragment can also be introduced into this Hamiltonian. This correlation potential is varied in a self-consistent manner to minimize the difference between the one-particle reduced density matrix (1-RDM) elements ($D_{pq}$) calculated at the mean-field level ($D^{\mathrm{LL}}_{pq}$) and the one calculated by the high-level ($D^{\mathrm{HL}}_{pq}$) method, thus improving the bath representation after each self-consistent cycle~\cite{5years_of_DMET}. The embedding Hamiltonian augmented with the chemical and correlation potentials can be represented as:
\begin{equation}
	\hat{H}_{\mathrm{emb}}=\sum_{pq}^{2N_A}\left(\tilde{h}_{pq}+u_{pq}\right)\hat{E}_{pq} + \frac{1}{2}\sum_{pqrs}^{2N_A}\tilde{V}_{pqrs}\hat{e}_{pqrs} - \sum_{p}\mu_{\mathrm{glob}}\hat{E}_{pp}
\end{equation}
A self-consistent DMET can be performed by solving for the correlation and chemical potentials such that the following error functions are minimized:
\begin{subequations}
	\begin{align}
		\Delta_{pq}(u)       & =D^{\mathrm{HL}}_{pq}(u)-D^{\mathrm{LL}}_{pq}  \\
		\Delta_N(\mu_{\mathrm{glob}}) & =N_{\mathrm{tot}}(\mu_{\mathrm{glob}})-N_{\mathrm{occ}}
	\end{align}
\end{subequations}

In most DMET calculations, the bath states obtained from the HF wave function ($u=0$) are used, and this is referred to as a ``one-shot DMET'' calculation. The total energy of the system can then be calculated using:
\begin{equation}
	E_{\mathrm{tot}}=E_{\mathrm{imp(=frag+bath)}}+E_{\mathrm{core}}+E_{\mathrm{nuc}}
\end{equation}

In periodic systems, electrons experience an effective periodic potential, and the electronic Hamiltonian commutes with the translation operator, which allows the electronic wave function to be expressed using Bloch functions. \citet{pham2019periodic} and \citet{cui2020efficient} utilized crystalline Gaussian-type orbitals (CGTOs), denoted by $\phi_{\mu}^{\mathbf{k}}(\mathbf{r})$, which are translational-symmetry adapted linear combinations of GTOs ($\bar{\phi}_{\sigma}(\mathbf{r})$), given by:

\begin{equation}
\phi_{\sigma}^{\mathbf{k}}(\mathbf{r}) = \sum_T  e^{i\mathbf{k}\cdot\mathbf{T}} \bar{\phi}_{\sigma}(\mathbf{r}-\mathbf{T})
\end{equation}

$\mathbf{T}$ is the translation vector, and $\mathbf{k}$ is the crystal momentum vector in the first Brillouin zone, which is generally sampled using uniform `k-points' mesh of size $N_k$. 

To adapt DMET to solid-state systems, the initial mean-field wave function is obtained by solving the periodic HF equation ($\hat{F}^{\mathbf{k}}\ket{\psi^{\mathbf{k}}}=\epsilon^{\mathbf{k}} \ket{\psi^{\mathbf{k}}}$) for each `k-point', followed by transforming the delocalized and dispersive crystalline orbitals into a localized basis for selecting the impurity subspace. The work by \citet{cui2020efficient} discusses several localization methods. One approach is to employ maximally localized Wannier functions \cite{marzari1997maximally,marzari2012maximally} (MLWFs) for the localization and construction of impurity subspace.

\begin{equation}
    \ket{\psi^{\mathbf{R}}} = \sum_{\mathbf{k}} U^{\mathbf{k}} \ket{\psi^{\mathbf{k}}} e^{-i\mathbf{k}\cdot \mathbf{R}}
    \label{eq:ktoR}
\end{equation}

The converged mean-field orbitals (${\psi^{\mathbf{k}}}$) in reciprocal space are transformed to real-space Wannier functions (${\psi^{\mathbf{R}}}$) using discrete Fourier transformation (Equation \eqref{eq:ktoR}). Localization to MLWFs is achieved via a unitary matrix $U^{\mathbf{k}}$, which is solved to minimize the spread functional $\Omega$ (= $\sum_n \bra{\psi_n^{\mathbf{R}}}\hat{\mathbf{r}}^2-\hat{\mathbf{r}}_n^2\ket{\psi_n^{\mathbf{R}}}$) for all the atoms in the unit cell, where $\mathbf{r}_n$ is the centroid of the Wannier function. Once the MLWFs are defined, the impurity subspace and the embedded Hamiltonian are constructed and the corresponding eigenvalue equation is solved in real space using the same procedure as that for the molecular system. On the other hand, if the self-consistent DMET procedure is employed, then the embedding model is constructed in real space while the mean-field wave function is iteratively updated in the momentum space~\cite{pham2019periodic}. This extension to treat solid-state systems has been termed as periodic DMET (pDMET). 

\begin{figure}[H]
	\centering
	\includegraphics[width=\textwidth]{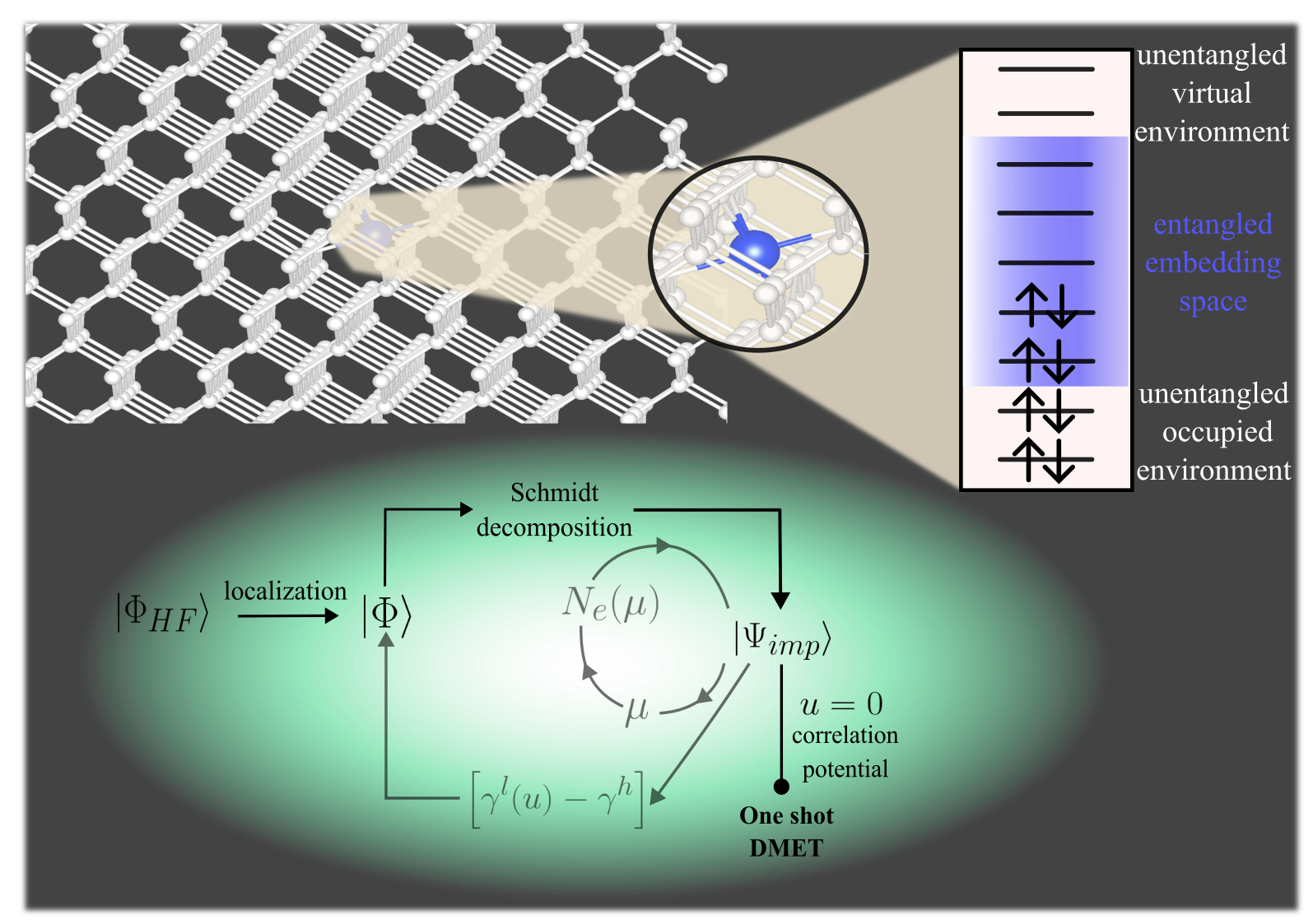}
	\caption{The general scheme of a DMET calculation involves identifying a region of localized chemical phenomenon, localization of the molecular orbitals, followed by a Schmidt decomposition to obtain the impurity wave function by selecting the DMET embedding space. This DMET impurity problem can be tackled with the high-level quantum chemistry methods of choice.}
	\label{fig:DMET-cat-2}
\end{figure}

Within this framework, one or more unit cells can serve as a fragment with the remaining cells forming the environment. This approach allows band structure calculations for simple systems such as periodic hydrogen chains and polyynes, utilizing the correlation potential \cite{pham2019periodic}. However, most pDMET applications centered on crystals with vacancies remain focused on gamma-point solutions (i.e. $N_k=1$) for the calculation of 
defect-localized vertical excitation energies (VEEs). \Cref{fig:DMET-cat-2} illustrates the general DMET workflow. 

Since its introduction, DMET has been applied to various condensed phase model systems~\cite{DMET_2DHubbard, DMET_Hubbard_Holstein, pDMET-1D,zheng2016ground, tsuchimochi2015density, fan2015cluster}. Beyond static DMET, efforts have been made to extend the method to non-equilibrium electron dynamics in strongly correlated systems. This approach, known as ``real-time DMET'', has been tested on single impurity Anderson models~\cite{kretchmer2018real}. Energy-weighted DMET was developed to account for quantum fluctuations from the fragment into the environment in a self-consistent manner~\cite{fertitta2018rigorous,fertitta2019energy}. DMET gradients were developed to describe multi-site reaction dynamics and were applied to describe proton transport in a small water cluster~\cite{li2023multi}. Additionally, DMET was extended to coupled fermion-boson systems in the one-dimensional Hubbard-Holstein model~\cite{sandhoefer2016density,doi:10.1021/acs.jctc.8b01116}. DMET has also been formulated for finite temperature by embedding a mean-field finite-temperature density matrix up to a given order in the Hamiltonian~\cite{sun2020finite}. DMET has also laid the foundation for more advanced embedding schemes. One such development is bootstrap embedding that overcomes DMET’s limitation of partitioning systems into non-overlapping fragments, enabling more flexible fragmentation~\cite{Welborn2016}. This approach has been successfully applied to molecules, surfaces, and solids~\cite{BE_mol,BE_largemol,meitei2023periodic}. 

The following sections provide a brief overview of the efforts in single-reference embedding applications with DMET, and a more detailed description of the utilization of DMET for multireference systems.

\subsection{Single-reference embedding with DMET}
Systems that are predominantly single-configuration can be described starting from a HF wave function, followed by M{\o}ller-Plesset (MP) perturbation theory \cite{MP1934, MPreview1, MPreview2} or coupled cluster \cite{Shavitt2009, Bartlett2007,iek1971,ek1966}, for quantitative accuracy. However, their application is constrained by polynomial scaling with system size ($N$). Second-order MP theory (MP2) scales as $\mathcal{O}(N^5)$, while CCSD and CCSD(T) scale as $\mathcal{O}(N^6)$ and $\mathcal{O}(N^7)$, respectively \cite{crawford2007introduction}. Various approaches have been developed to mitigate these scaling challenges, including integral decomposition \cite{feyereisen1993use}, natural orbital-based truncation schemes \cite{taube2005frozen}, localization-based methods \cite{saebo1993local, neese2009efficient}, and embedding-based methods~\cite{knizia2012density, bulik2014electron, wouters2016}. 

\begin{figure}[H]
    \centering
\includegraphics[width=\textwidth]{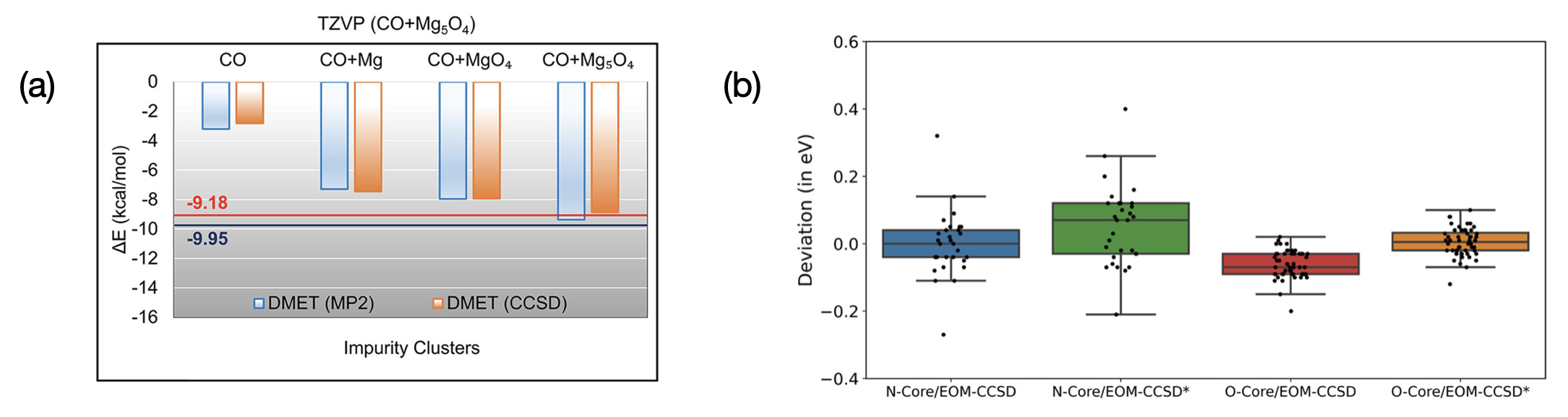}
    \caption{(a) Binding energy of the \ce{CO} adsorption on the \ce{MgO} surface. The deviation of the DMET-MP2/CCSD methods from their non-embedded version for different embedded fragment size~\cite{Mitra2022}. (b) Core ionization computed using DMET-CVS-EOM-CCSD. The deviation of the embedded version from its non-embedded counterpart for the 82  N- and O K-edge core ionization calculated at EOM-CCSD/* level, where * denotes the perturbative triples corrections on top of the EOM-CCSD~\cite{jangid2024core}. Figures reproduced with permission from references \citenum{Mitra2022,jangid2024core}. }
    \label{fig:srembd}
\end{figure}

Since an embedding method like DMET makes it possible to treat only a subset of the system with these correlated wave function methods, one only needs to solve the impurity problem to obtain $E_{\mathrm{imp}}$ with the expensive correlated methods.
The size of the problem under study is reduced from $N$ to $N_{\mathrm{imp}}$. For instance, Cui et. al. investigated the ground-state electronic properties of nickel oxide~\cite{cui2020efficient} and magnetic states of the cuprate superconductors, both in their doped and pristine states \cite{Cui2022} using periodic DMET with CCSD as high-level solver. Furthermore, DMET has also emerged as a promising alternative for accurately modeling weak interactions such as van der Waals forces, relevant to heterogeneous catalysis in extended systems, without relying on cluster models or QM/MM approaches with ambiguous boundaries~\cite{Boese2013}. In this direction, \citeauthor{Lau2021} presented a slightly modified version of DMET, and used that framework with MP2/CCSD and its perturbative-triples versions to model the adsorption of water molecule on \ce{LiH}, hBN and graphene surfaces. Another study on the adsorption of \ce{CO} molecule on the \ce{MgO}(100) surface, used pDMET with MP2/CCSD solvers to obtain binding energies and showed excellent agreement between the embedding methods and their non-embedding equivalents~\cite{Mitra2022}. The convergence of the binding energy with embedding size is shown in \Cref{fig:srembd}(a). It was found that in the case of \ce{CO}@\ce{MgO} adsorption, it is not sufficient to use smaller DMET fragments \ce{CO} or \ce{CO}+\ce{Mg}, rather a larger fragment \ce{CO}+\ce{MgO_4} is required, necessitating either large memory or high I/O for the 4-index electron repulsion integrals. \citeauthor{Mitra2022} utilized density fitting to reduce the memory requirements by 5-6 folds for the \ce{CO}@\ce{MgO} case.

In core-level spectroscopy, the core orbitals are usually highly localized, and core holes are primarily influenced by neighboring atoms, which makes embedding methods ideal for accurately modeling core-ionization processes. The core-valence separated-EOM-CCSD \cite{coriani2015communication} with a triple-$\zeta$ basis set reliably produces ionization energies within $<1$ eV of the experimental values, but these methods have high scaling ($\mathcal{O}(N^6)$ or higher)  and are limited to model systems. \citet{jangid2024core} combined the DMET framework with CVS-EOM-CCSD to compute core-ionization energies, achieving a mean absolute error of 0.06 eV across 82 N- and O K-edge core states in amino acids (shown in Figure \ref{fig:srembd}(b)). This approach significantly reduced the computational cost while maintaining high accuracy, enabling applications to larger systems such as uracil-hexamer, aza-fullerene, and chlorophyll $a$. \citet{lau2024optical} investigated  different flavors of orbitals, such as canonical, localized and natural transition orbitals for excitation energy EOM-CCSD computations of point-defects in extended systems. 

An advancement by \citeauthor{Nusspickel2022}, building on DMET, introduces an alternative to fragment-size scaling: an expanded bath space constructed from additional states inspired by the pair natural orbital approach~\cite{Nusspickel2022}. This new, controllable parameter enables systematic improvements in material property predictions, such as correlation energies in diamond, graphene, and \ce{SrTiO3}. In another study, approaches for reconstruction of nonlocal observables like total energy on fragmented solutions starting from DMET as the embedding scheme have been described~\cite{nusspickel2023effective}.

\subsection{Multireference embedding with DMET}
Often, a single electronic configuration cannot adequately describe a fragment’s electronic structure due to multiple nearly degenerate states (known as static correlation). This is common in bond breaking, transition-metal complexes, and point defects in solids, where localized electronic states within the band gap may arise. Excited states can also be defect-localized and multiconfigurational in nature, as seen in color centers in diamond and silicon carbide \cite{bockstedte2018ab}, where single-reference methods fail.

In such cases, multireference methods are essential to tackle near degeneracies of electronic states. CASSCF \cite{RoosComplete1980, RoosComplete1987} accounts for static correlation by performing a full CI within a selected set of orbitals, referred to as the active space. systems where
CASSCF becomes computationally prohibitive, variants like the restricted active space SCF (RASSCF) \cite{Malmqvistrestricted1990} or generalized active space (GASSCF) \cite{Fleiggeneralized2001,Mageneralized2011} reduce the cost by limiting the CI expansion in a systematic way.

While CASSCF provides qualitatively accurate wavefunctions, it neglects correlation effects outside the active space (primarily dynamical electron correlation). Enlarging the active space can recover more external, dynamic correlation, but becomes intractable for large systems and active spaces. Post-SCF methods address this. Multireference many-body perturbation theories (MRPT) methods like CAS second-order perturbation theory CASPT2 \cite{AnderssonSecond1990, AnderssonSecond1992} or $N$-electron valence-state second-order perturbation theory (NEVPT2) \cite{AngeliIntroduction2001} are widely used to recover dynamic correlation on top of a CASSCF reference, although they can be computationally demanding. Alternatively, multireference adaptations of traditionally single-reference methods -- such as multireference configuration interaction (MRCI) \cite{BuenkerIndividualized1974,BuenkerApplicability1978} and multireference coupled cluster (MRCC) \cite{IvanovMultireference2012} -- have been developed, but these approaches also come with high computational cost, as they require a preceding multiconfigurational SCF calculation followed by CI or CC treatment. More recently, multireference algebraic diagrammatic construction (MR-ADC) methods \cite{sokolov2018multi,mazin2021multireference} have emerged as promising alternatives, offering lower computational cost compared to more traditional multireference techniques while still aiming to capture missing dynamic correlation effectively. 

Gagliardi, Truhlar, and co-workers\cite{LiManni2014, ghoshCombining2018, zhouElectronic2022} have developed a post-SCF method that combines wave function theory with density functional theory, known as multiconfiguration pair-density functional theory (MC-PDFT). MC-PDFT offers accuracy comparable to traditional multireference perturbation theory methods, but at significantly reduced computational cost. In MC-PDFT the total energy is the sum of the classical energy (kinetic energy, nuclear attraction, and classical Coulomb energy of the electrons) calculated from a multiconfigurational wave function and the nonclassical energy (often called the exchange–correlation energy) obtained using a functional of the total density ($\rho$) and the on-top pair density ($\Pi$), referred to as the on-top functional $E_{\mathrm{ot}}$.

\begin{equation}
\label{eq:pdft_energy}
	E^{\mathrm{PDFT}} = \sum_{pq} h_{pq}D_{pq} + \frac{1}{2}\sum_{pqrs}V_{pqrs}D_{pq}D_{rs} + E_{\mathrm{ot}}\bqty{\rho, \Pi}
\end{equation}
MC-PDFT has been shown to be more accurate than CASSCF and as accurate as CASPT2 in predicting vertical excitations on a small subset of important organic chromophores \cite{hoyerMulticonfiguration2016}, and as accurate as NEVPT2 in predicting VEEs for the QUESTDB data set (containing over 500 vertical excitations from small and medium-sized main-group molecules)~\cite{kingLargeScale2022}. It has been used in combination with different types of MR wave functions, including GASSCF \cite{odoh2016separated,ghosh2017generalized} and DMRG~\cite{sharma2019density}. The advantage of MC-PDFT is that it scales similarly to CASSCF and is therefore significantly faster and more affordable than MRPT methods. 

To capture dynamic correlation, the strongly-contracted NEVPT2 version \cite{Angeli2002} as available in PySCF \cite{PySCF0,PySCF1}, scales as $\mathcal{O}(N_{\mathrm{core}}^2N_{\mathrm{vir}}^2)$~\cite{Anderson2020}, with $N_{\mathrm{core}}$ core and $N_{\mathrm{vir}}$ virtual orbitals. In NEVPT2-DMET, where $N_{\mathrm{core}}$ and $N_{\mathrm{vir}}$ are significantly smaller, this becomes more affordable~\cite{mitra2021excited}. Still, NEVPT2 requires the full 3-RDM and scales as $\mathcal{O}(N_{\mathrm{act}}^6)$ in storage. MC-PDFT, needing only 1- and 2-RDMs, offers a lower-cost alternative. Feeding CAS-DMET RDMs into Equation \eqref{eq:pdft_energy} defines the DME-PDFT method \cite{Mitra2023,Verma2025dmepdft}, which captures correlation beyond the embedding space.

CAS-DMET, with or without bath truncation, has been applied to \ce{N-N} bond dissociation in azomethane and pentyl diazine \cite{pham2018can}. It agrees well with CASSCF near equilibrium and—with bath truncation—also at dissociation. DME-PDFT is more robust to truncation, as it treats the full system~\cite{Verma2025dmepdft}.

DMET has also been extended to include spin–orbit coupling (SOC) effects. CAS-DMET combined with state interaction (CASSI-SO) \cite{ai2022efficient} yields accurate zero-field splitting (ZFS) for 3$d$ transition-metal single-ion magnets. Regularized direct inversion of iterative subspace (R-DIIS) accelerates convergence of low-level DMET restricted open-shell HF solutions. This has been expanded to NEVPT2-DMET \cite{guan2025density} and adapted for 4$f$ systems~\cite{ai2025density}. In parallel, one-shot DME-PDFT has also been applied to spin-state energetics in extended transition metal systems like \ce{Fe[N(H)Ar^*]_2} and \ce{Ni[C_{90}C_{20}H_{120}]^{2+}}. DME-PDFT and NEVPT2-DMET outperform truncated-model calculations predicting VEEs, regardless of the complexity or nature of the surrounding ligand environment~\cite{Verma2025dmepdft}. Collectively, these advancements highlight the versatility of DMET-based embedding for accurate and scalable multireference treatments in molecular systems.

A significant development was the extension of these multireference DMET frameworks to solid-state systems~\cite{pham2019periodic}, enabling accurate modeling of multiconfigurational states arising from localized defects or adsorbates. In particular, accurately computing VEEs in defects is crucial for the design of materials relevant to quantum technologies~\cite{Muechler2022,repa2023lessons,somjit2025nv,onizhuk2025colloquium,jin2025first}. Periodic CAS-DMET and post-CAS-DMET define impurity subspaces within an infinite system, with the embedding subspace encompassing the defect site.

\begin{figure}[H]
	\centering
	\includegraphics[width=\textwidth]{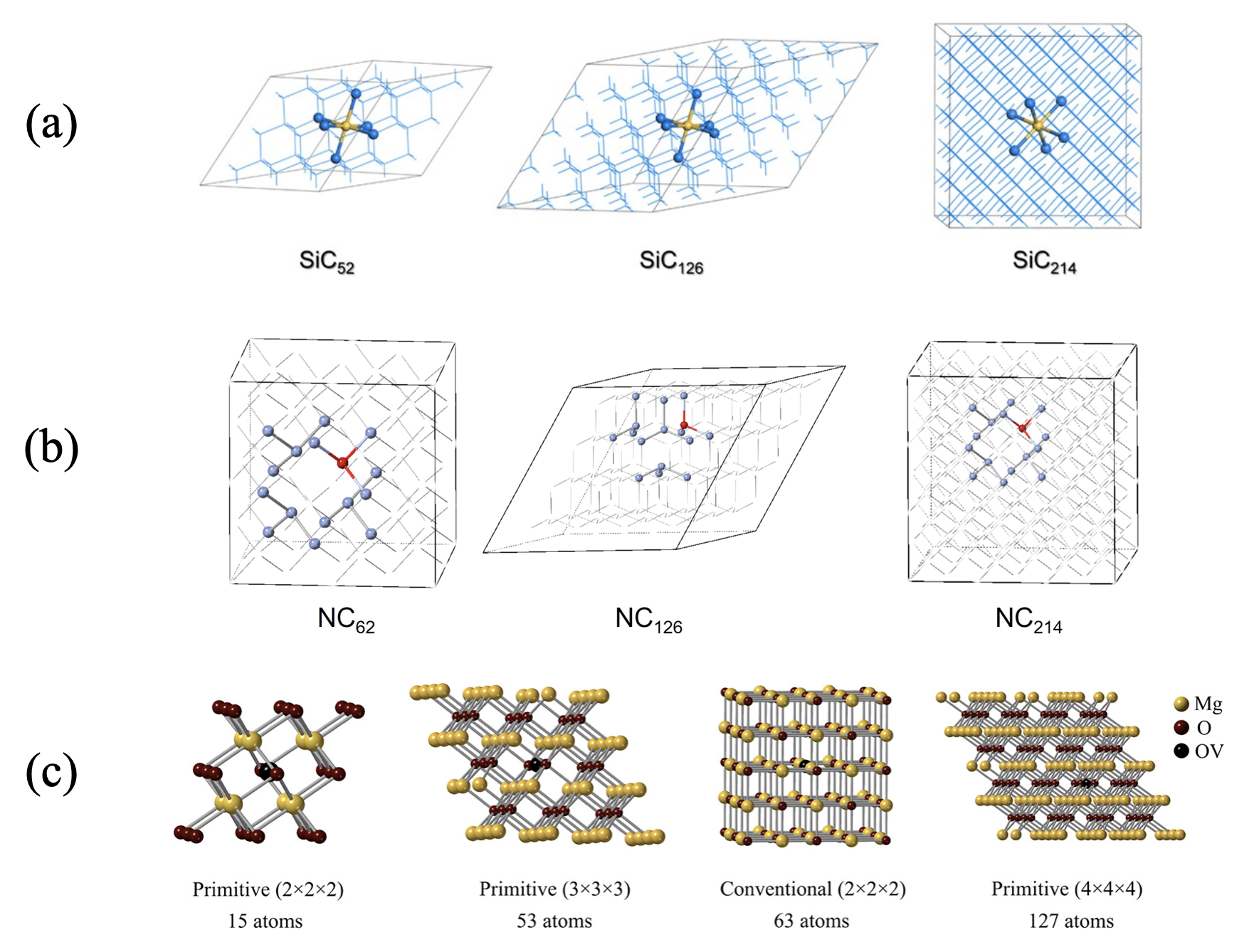}
	\caption{The periodic supercells studied through CAS-DMET and post-CAS-DMET methods. (a) Three supercell models: \ce{SiC54}, \ce{SiC126}, and \ce{SiC214} for \ce{SiV^0} defect in diamond. Figure reproduced with permission from reference \citenum{mitra2021excited}. (b) Three supercell models: \ce{NC62}, \ce{NC126}, and \ce{NC214} for \ce{NV-} defect in diamond. In each of them the \ce{NC15} impurity cluster for the DMET calculations has been highlighted. Figure reproduced with permission from reference \citenum{haldar2023local}. (c) Four different supercells employed in the periodic DMET study along with their nomenclature and the respective number of atoms in each. Figure reproduced with permission from reference \citenum{mitra2021excited,haldar2023local,Verma2023}.}
    \label{fig:all-supercells-for-pdmet}
\end{figure}

\begin{figure}[H]
    \centering
    \includegraphics[width=\textwidth]{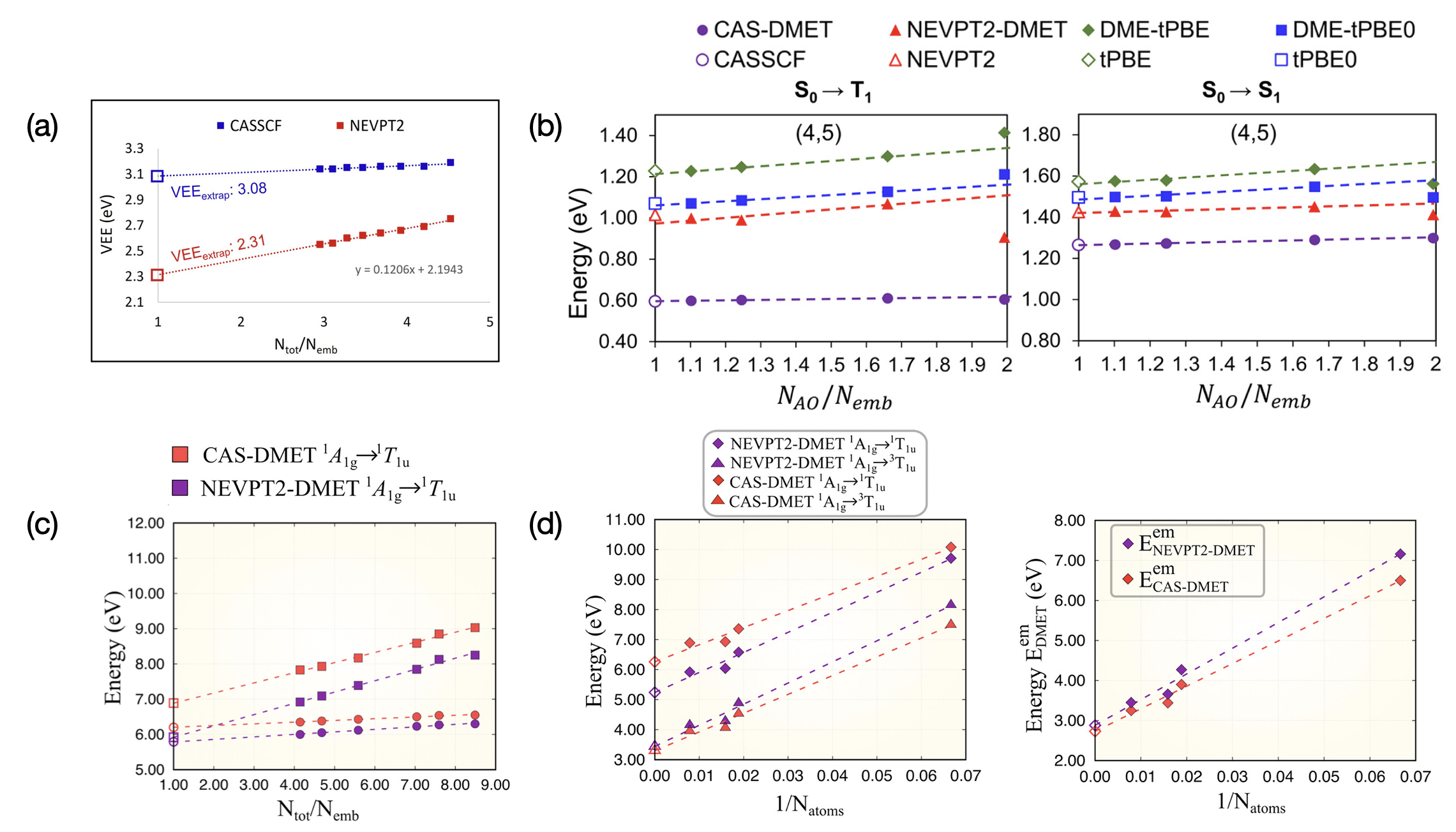}
    \caption{Linear extrapolation to non-embedding in (a) the 214 atom-containing supercell of \ce{NV-} studied with CAS-DMET and NEVPT2-DMET~\cite{haldar2023local}, (b) oxygen divacancy on \ce{MgO}(100) surface (\ce{Mg_{18}O_{18}}) studied with CAS-DMET, NEVPT2-DMET, DME-tPBE and DME-tPBE0~\cite{Mitra2023}. (c) Linear extrapolation to the non-embedding (primitive ($4 \times 4 \times 4$) supercell) and thermodynamic limits in oxygen vacancy in bulk of \ce{MgO} studied with CAS-DMET and NEVPT2-DMET~\cite{Verma2023}. Figure reproduced with permission from references \citenum{haldar2023local,Mitra2023,Verma2023}.}
    \label{fig:extrapol-emb}
\end{figure}

These methods have been applied to compute VEEs for defects such as \ce{NV^-} \cite{haldar2023local} and \ce{SiV^0} \cite{mitra2021excited} in diamond, and neutral oxygen vacancies (\ce{OV^0}) in bulk and surface MgO. \Cref{fig:all-supercells-for-pdmet} shows the defects and supercells studied using periodic DMET. To predict VEEs for a given defect, one generates increasingly large supercells and selects maximally localized Wannier functions centered on the defect as DMET fragments. CAS-DMET and NEVPT2-DMET VEEs for each supercell are then plotted against the ratio of total to embedding basis functions. For example, \Cref{fig:extrapol-emb}(a) from reference \citenum{haldar2023local} shows extrapolation to the non-embedding limit for a 214-atom supercell of \ce{NV^-}. While CAS-DMET VEEs remain invariant with embedding size, NEVPT2-DMET energies vary linearly with the inverse embedding size. Though CAS-DMET and NEVPT2-DMET have been applied to various bulk and surface defects, DME-PDFT has so far only been tested on oxygen mono- and di-vacancies on \ce{MgO}~\cite{Mitra2023}. \Cref{fig:extrapol-emb}(b) shows similar extrapolation for an \ce{OV} on \ce{MgO(100)} using periodic DME-PDFT, which shows low or comparable sensitivity to embedding size, making it a more efficient alternative to NEVPT2-DMET. These early results highlight DME-PDFT’s potential as a scalable, cost-effective multireference method for extended systems.

The VEEs above were computed only at optimized ground-state geometries. For bulk \ce{OV^0} in \ce{MgO}, CAS-DMET/NEVPT2-DMET VEEs were renormalized using GW-time-dependent DFT (TD-DFT) optimized excited-state geometries \cite{jin2023excited} and this opened the direction of exploring the emission mechanism~\cite{Verma2023}. This approach can be extended to other defects requiring adiabatic excitation energies. In addition to linear extrapolation to the non-embedding limit, \Cref{fig:extrapol-emb}(d) shows a second extrapolation to the thermodynamic limit based on inverse supercell size.

The electronic properties of an \ce{Fe} impurity in \ce{AlN} were recently studied using CAS-DMET and NEVPT2-DMET~\cite{otis2025strongly}. This work examined the convergence of VEEs with respect to active space size and compared DMET results to those from QDET. Spin-flip TD-DFT was used to optimize excited-state geometries, enabling comparison of DMET excitation energies with experimental zero-phonon line data.

All excitation energies computed using CAS-DMET and NEVPT2-DMET for the studied defects are summarized in \Cref{tab:dmet-periodic-applications}. These methods are expected to be increasingly applied to diverse point defects in the search for useful qubits and materials for quantum technologies.

\begin{table}[H]
	\centering
	\begin{tabular}{cccccc}
		\hline                                                                                               \\
		System                  & Excitation                     & CAS- & NEVPT2-  & Experimental/ Reference\\
		                        &                                & DMET (eV) & DMET (eV)      &     (eV)          \\

		\hline\\
		\ce{F_S^0}@\ce{MgO} (110)       & $S_0\rightarrow S_1$           & 3.77 & 3.62     &1-5\cite{henrich1980energy,kramer2002mechanism,Wu1992}                       \\
		                        & $S_0\rightarrow T_1$           & 2.35 & 2.28       &-                     \\
		\ce{SiV^0}                 & $T_0 \rightarrow$ $T_1$        & 2.26 & 2.39      &-                     \\
		                        & $T_0 \rightarrow$ $T_2$        & 2.44 & 2.47      &1.31 \cite{green2019electronic}                     \\
		                        & $T_0 \rightarrow$ $T_3$        & 2.44 & 2.46      &1.31 \cite{green2019electronic}                     \\
		                        & $T_0 \rightarrow$ $T_4$        & 3.16 & 2.61     &1.79\cite{Ma2020}                       \\
		                        & $T_0 \rightarrow$ $S_1$        & 0.50 & 0.51      &0.34\cite{Ma2020}                    \\
		                        & $T_0 \rightarrow$ $S_2$        & 0.50 & 0.51      &0.34 \cite{Ma2020}                     \\
		                        & $T_0 \rightarrow$ $S_3$        & 1.36 & 1.14    &0.58                       \\
		\ce{F^0} @MgO              & $^1A_{1g}\rightarrow ^1T_{1u}$ & 6.26 & 5.24   &5.03 \cite{chen1969defect}                      \\
		                        & $^3T_{1u}\rightarrow ^1A_{1g}$ & 2.74 & 2.89      &2.40\cite{rosenblatt1989luminescence}                      \\
		\ce{NV^-} & $^3A_2 \rightarrow$ $^1E$      &   0.85   & 0.50             & 0.50-0.59  \cite{goldman2017erratum}   \\
		                        & $^3A_2 \rightarrow$ $^1A_1$    &   2.96   & 1.56             & 1.76-1.85\cite{goldman2017erratum}     \\
		                        & $^3A_2 \rightarrow$ $^3E$      &  3.08    & 2.31              & $\sim$2.18 \cite{davies1976optical}   \\
		                        & $^1E \rightarrow$ $^1A_1$      &  2.09    & 1.02              & $\sim$1.26 \cite{kehayias2013infrared}   \\
                                \ce{Fe^0_{Al}}@ AlN& $^4E\rightarrow^6A_1 $&1.94&1.46&1.30\cite{baur1994determination}\\
                                \\

		\hline
	\end{tabular}
	\caption{Vertical excitation energies for various periodic systems studied with CAS-DMET and NEVPT2-DMET~\cite{mitra2021excited, haldar2023local,Verma2023}.}
	\label{tab:dmet-periodic-applications}
\end{table}

We conclude this section by noting the extensive progress in the development of quantum embedding methods, especially within the context of density matrix embedding theory. These approaches are increasingly being applied to realistic systems, both molecular and periodic, beyond simplified models, as shown by the examples described above, and they are beginning to incorporate the essential physics required to capture their complexity, including strong electron correlation. Given the growing number of available embedding methods, a key priority for future work is the systematic comparison of these approaches to assess their relative strengths and limitations across different classes of systems. Such benchmarking efforts will be essential to identify which embedding strategies are most suitable for which types of problems.

Key future directions for advancing the field include developing quantum embedding methods to describe core excitations in multireference systems \cite{fouda2025computation}, the dynamics of extended systems, incorporating spin–phonon interactions, and enabling geometric relaxation. Furthermore, quantum embedding-based approaches hold great promise for training machine-learned force fields that can facilitate the efficient study of dynamic processes in complex systems.

\section{Fragment-based multireference quantum algorithms} 
\begin{figure}
    \centering
    \includegraphics[width=\linewidth]{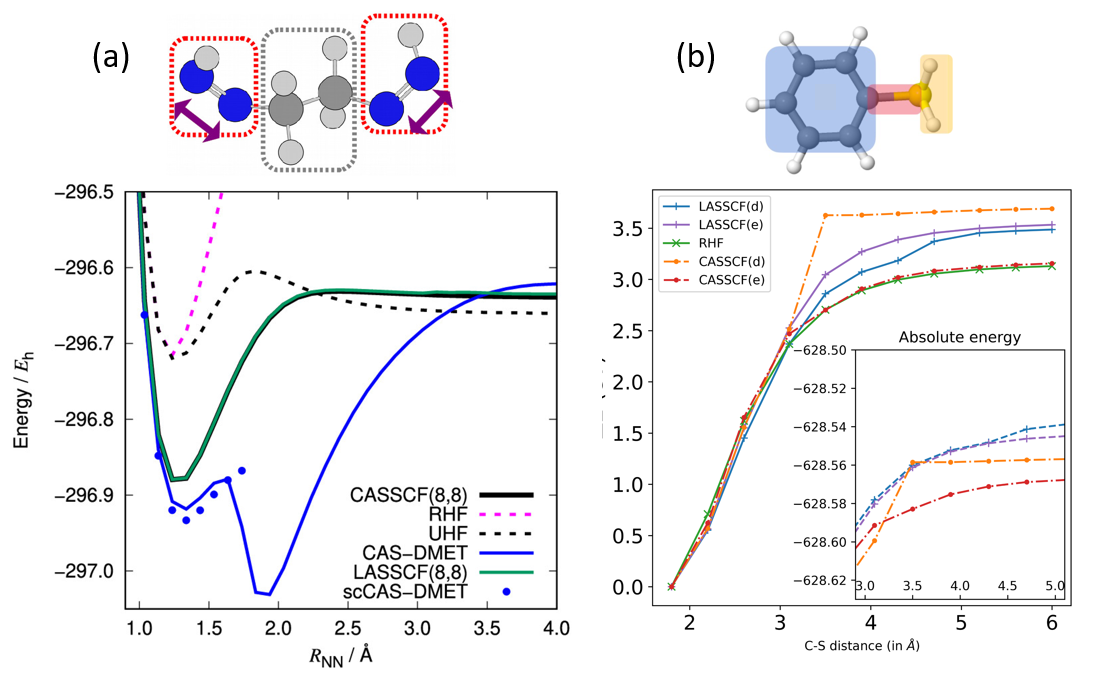}
    \caption{\textit{left}: Bisdiazing bond stretching potential energy curves with various methods. LASSCF reproduces CASSCF faithfully while multifragment DMET is incorrect.  \textit{right}: Ground state, heterolytic dissociation potential energy scan of monophenylsulfonium cation, inset shows the absolute energies. Energy is plotted relative to that at the equilibrium geometry in all cases, and ``(d)'' (dissociation) indicates a scan initiated from dissociation, while ``(e)'' (equilibrium) indicates a scan initialized from the equilibrium geometry. Figure reproduced with permission from reference \citenum{wang2024}.}
    \label{fig:las-sulfonium}
\end{figure}

While multireference methods integrated with the DMET framework in the context of single impurities offer substantial cost reduction, this does not provide any significant advantage in accessing larger active spaces. One way to address this issue is by creating multiple fragments when electron correlation is localized in multiple regions of a system.
In scenarios requiring multiple embedding spaces -- such as the case of two fragments or a bond breaking, CAS-DMET fails~\cite{pham2018can}, primarily due to the use of ambiguous error functions to match low-level and high-level density matrices~\cite{hermes2019multiconfigurational}. The localized active space self-consistent field (LASSCF) \cite{hermes2019multiconfigurational,hermes2020variational} method, also known as the cluster mean-field (cMF) method \cite{cmf,papastathopoulos2023symmetry}, creates unentangled active spaces with each having its own bath made through Schmidt decomposition. The LASSCF wave function is given by:
\begin{equation}
	\ket{\Psi_{\text{LAS}}} = \bigwedge_{K} (\ket{\Psi_{A_K}}) \wedge \ket{\Phi_{\text{SD}}}
\end{equation}

where \( \ket{\Psi_{A_K}} \) denotes the many-body (generally FCI) wave function of the \( K \)th localized subspace, and \( \ket{\Phi_{\text{SD}} }\) denotes the single-determinantal wave function of closed-shell occupied inactive orbitals. The energy of the system (\( E_{\text{LAS}} \)) is obtained through variational optimization and is expressed as:
\begin{equation}
	E_{\text{LAS}} = \langle \Psi_{\text{LAS}} | \hat{H} | \Psi_{\text{LAS}} \rangle
\end{equation}
where \( \hat{H} \) denotes the molecular Hamiltonian (Equation \eqref{eq:mol-ham}). LASSCF was shown to reproduce the corresponding CASSCF reference potential energy curve for stretching a bisdiazine molecule, whereas one-shot CAS-DMET and self-consistent CAS-DMET do not yield physically meaningful curves. A recent study has also demonstrated that LASSCF provides a more consistent orbital evolution along the dissociation potential energy surface of sulfonium salts compared to CASSCF, as shown in \Cref{fig:las-sulfonium}(b). It accurately predicts homolytic and heterolytic dissociation pathways while maintaining relative energies comparable to CASSCF. Additionally, LASSCF has been shown to reliably predict spin-state energy gaps when fragments are well separated~\cite{hermes2019multiconfigurational,hermes2020variational,riddhish_spin}.

However, LASSCF can fall short in cases where the assumption of mean-field interaction is too drastic~\cite{pandharkar2022localized, agarawal2024automatic}. In a classical computing framework, the interfragment interaction energy can be computed using state interaction (LASSI)\cite{pandharkar2022localized, agarawal2024automatic,hermes2025localized} for recovering interfragment correlations and MC-PDFT (Equation \eqref{eq:pdft_energy}) to address dynamic correlation \cite{Pandharkar2020,hermes2025localized} if the underlying LASSCF/LASSI wave function is qualitatively correct. The cost of LASSI can suppress CASSCF scaling for a few fragments, but rises steeply and reintroduces factorial scaling with the number of fragments, and consequently, the size of active space, characteristic of CASSCF. The FragPT2 method recently introduced by \citet{koridon2025fragpt2} recovers inter-fragment correlation between localized fragments using multireference perturbation theory.

Quantum computers offer an alternative approach by providing theoretical exponential speedups. The VQE is a popular hybrid quantum-classical algorithm that delegates the classically intractable component of the computation to the quantum device, specifically state preparation and energy measurement, with the wave function parameters being classically optimized. Current hardware limitations on qubit numbers and circuit depth mean that for algorithm development on noisy intermediate-scale quantum (NISQ) hardware, active-space sizes must be limited. To address this limitation, VQE has been utilized as a high-level solver for projection-based WF-in-DFT embedding,\cite{manby2012, Rossmannek2023} and classical DMET has been adapted as a hybrid classical-quantum algorithm where the embedding Hamiltonian problem is solved using a quantum simulator, and the density matrix matching is performed on classical hardware~\cite{Cao2023,erakovic_high_2025}. Classical fragmentation schemes have been combined with quantum algorithms, such as in the FMO/VQE algorithm where fragmentation is performed using the fragment molecular orbital (FMO) approach,\cite{lim_fragment_2024} and the divide-and-conquer unitary coupled cluster algorithm (DC-qUCC/VQE) where the linear-scaling classical DC scheme is used for fragmentation~\cite{yoshikawa_quantum_2022}. Classical bootstrap embedding has also been adapted to quantum hardware where the density matrices of the fragment boundaries are further matched using SWAP gates on a quantum simulator~\cite{bootstrap,erakovic_high_2025}. The many-body expansion (MBE) has also been combined with the VQE~\cite{ma_divide-and-conquer_2022}. In addition, efforts have been made to reduce the quantum resources needed for high-level VQE solvers for DMET problems using the energy sorting strategy~\cite{fan2023circuit,DMET-VQE}. All these methods involve performing a quantum algorithm in fragments to accommodate current hardware limitations. 

\begin{figure}[H]
	\centering
	\includegraphics[width=\textwidth]{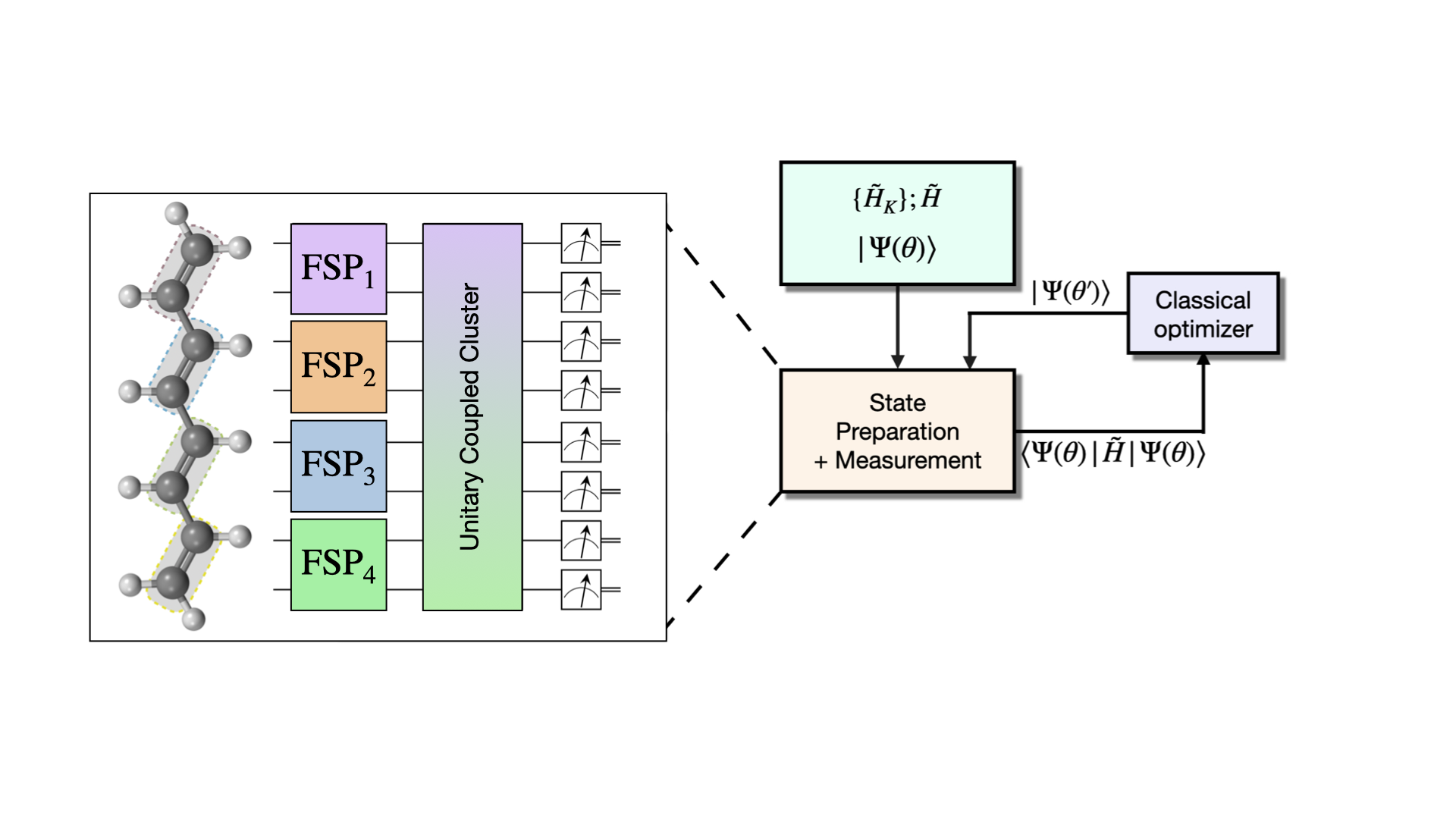}
	\caption{ Schematic of a general LAS-UCC algorithm. The system of interest is first separated into distinct fragments, with \{$\hat{H}_K$\} representing the Hamiltonian for fragment $K$, and $\hat{H}$ representing the Hamiltonian for the full active space. A fragment state preparation (FSP) method is used to solve and load each fragment's wave function, based on a classical LASSCF calculation. Correlation between fragments is then added variationally through a UCC ansatz, with the circuit being prepared and the energy $E(\theta) = \expval{\hat{H}}{\Psi(\theta)}$ measured during a single VQE iteration.}
	\label{fig:LAS-UCC}
\end{figure}

 A philosophically different approach to combining classical fragmentation and quantum algorithms is the localized active space unitary coupled cluster method (LAS-UCC) that employs the unitary coupled cluster singles and doubles (UCCSD) ansatz \cite{VQE,taube_new_2006} to effectively capture electron correlation between fragments using VQE~\cite{Otten2022, dcunha_state_2024}. In LAS-UCC, the LASSCF state vector for each fragment can be loaded onto the qubit register either using direct initialization (DI) or QPE, and the correlation between fragments can be recovered using a standard VQE procedure as shown in \Cref{fig:LAS-UCC}. The UCC ansatz:
\begin{align}
    |\Psi_{\mathrm{UCC}}\rangle &= e^{\hat{T}- \hat{T}^{\dagger}} |\Psi_{\mathrm{ref}}\rangle \\
    \hat{T} &= \hat{T}_1 + \hat{T}_2 + \dots
\end{align}
is used as a physically-motivated ansatz with generalized single ($\hat{T}_1$) and double ($\hat{T}_2$) excitations since the reference wave function, $|\Psi_{\mathrm{ref}}\rangle$, is now the multireference LASSCF wave function. The purpose of initializing the VQE with LASSCF is to provide an efficient way to recover missing correlation between fragments -- which would be too expensive to do classically with a method such as state interaction -- while also providing the VQE with a trial wave function that has a larger overlap with the target ground state.

\begin{figure}[H]
	\centering
	\includegraphics{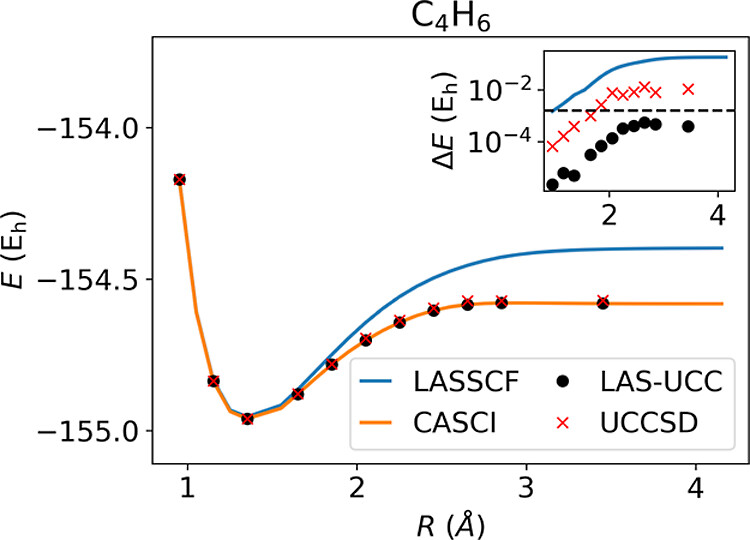}
	\caption{ Energies for \textit{s-trans}-butadiene calculated by CASCI, LASSCF, and LAS-UCC. The inset shows the error of LASSCF and LAS-UCC with respect to CASCI. The black dashed line represents the chemical accuracy. LAS-UCC obtains chemical accuracy across the potential energy surface, whereas LASSCF, that cannot accurately represent the correlation between the fragments, fails to obtain chemical accuracy for most points. Reproduced with permission from reference \citenum{Otten2022}.}
	\label{fig:LAS-UCC-c4h6}
\end{figure}

LAS-UCC singles and doubles (LAS-UCCSD) has shown promising results for model systems like strongly correlated geometries of \ce{H4} and \textit{s-trans}-butadiene as well as more challenging multimetallic molecules like  \ce{[Cr_2(OH)_3(NH_3)_6]^{3+}} and \ce{[Mn(NH3)4{oxamide}][Cu(NH3)2]^{2+}}. When the two methylene units are considered as LAS fragments and the double bonds within each fragment are stretched, the interfragment correlation increases and a double bond forms between the fragments (\Cref{fig:LAS-UCC-c4h6})\cite{Otten2022}. In this case study, LASSCF diverges significantly from the corresponding complete active space configuration interaction (CASCI) limit, whereas LAS-UCCSD captures most of the interfragment correlation.

The bimetallic compound \ce{[Cr_2(OH)_3(NH_3)_6]^{3+}} represents an interesting case. This system models the Kremer's dimer system that has been studied extensively in the literature~\cite{bennie_electronic_2012,pantazis_meeting_2019,sharma_magnetic_2020}. Various studies have suggested that a large active space of (30,22), including ligand orbitals is required to model the superexchange processes between the Cr centers\cite{pantazis_meeting_2019,sharma_magnetic_2020,hermes2025localized} and to correctly predict the value of $J$, the magnetic coupling constant, expressed as the difference between the high spin and low spin energies, scaled by the difference in $\langle\hat{\mathbf{S}}^2\rangle$ values of the states:
\begin{equation}
J=-\frac{E_{\mathrm{HS}}-E_{\mathrm{LS}}}{\langle\hat{\mathbf{S}}^2\rangle_{\mathrm{HS}}-\langle\hat{\mathbf{S}}^2\rangle_{\mathrm{LS}}}
\end{equation}

\begin{table}
    \centering
    \begin{tabular}{l|c|r}
        \hline
        Method & Active Space & $J$ (cm$^{-1}$) \\
        \hline
        \hline
        CASSCF\cite{MitraEtAl2024}& (6,6) &  -13.8\\
        CASCI@LAS\cite{MitraEtAl2024} & (6,6) &  -11.6\\
        LASSCF\cite{MitraEtAl2024} & ((3,3),(3,3)) & 11.2 \\
        LAS-UCCSD\cite{MitraEtAl2024} & ((3,3),(3,3)) & -11.6 \\
        LAS-USCCSD\cite{MitraEtAl2024} 5\% & ((3,3),(3,3)) & -8.3 \\
        LAS-QKSD\cite{dcunha_fragment-based_2024} & ((3,3),(3,3)) & -10.1 \\
        \hline
    \end{tabular}
    \caption{Magnetic coupling constant ($J$) values for \ce{[Cr_2(OH)_3(NH_3)_6]^{3+}} for various multireference methods and corresponding active spaces.}.
    \label{cr_dimer}
\end{table}
Results for this model system are presented in Table \ref{cr_dimer}. LASSCF qualitatively fails to predict the correct low-spin singlet ground state, with an incorrect sign of $J$. LAS-UCCSD reintroduces the significant interfragment correlation, thereby recovering both qualitative and quantitative accuracy in relation to the exact diagonalization (CASCI@LAS). However, a significant challenge in the practical implementation of the LAS-UCCSD method is the large number of UCCSD parameters for inter-fragment excitations required for energy optimization. 
Exploring and exploiting parameter redundancy to reduce circuit depths in the UCCSD ansatz is an emerging field of research~\cite{mehendale_exploring_2023, mullinax2025large, fleury_nonunitary_2024, fedorov_unitary_2022}.
A LAS-UCCSD calculation on \ce{[Cr_2(OH)_3(NH_3)_6]^{3+}} employing the def2-SVP basis for \ce{C}, \ce{N}, \ce{O}, and \ce{H} atoms, and the def2-TZVP basis for \ce{Cr} atoms, requires 774 parameters, which translates to the order of tens of thousands of quantum gates. 
To address this, the LAS unitary selected coupled cluster singles and doubles (LAS-USCCSD) method was developed \cite{MitraEtAl2024,mitra2024correction,verma2025polynomial}: a multireference implementation of the previously developed unitary selected coupled cluster algorithm \cite{fedorov_unitary_2022}, that leverages LAS-UCC energy gradients to classically preselect the most significant parameters for VQE calculations. This results in qualitative accuracy with a $J$ value of -8.3 cm$^{-1}$ while keeping just 5\% of parameters.
Other post-LAS methods include the use of quantum Krylov subspace diagonalization (QKSD) as the quantum algorithm to correlate the LASSCF fragments~\cite{dcunha_fragment-based_2024}. In this case, the difficult nonlinear optimization of the VQE is replaced by a subspace diagonalization using real-time evolved states. \Cref{cr_dimer} shows that results within 2 cm$^{-1}$ of the reference can be obtained using this subspace-based algorithm.

The performance of VQE algorithms, in general, relies on the parameterized wave function ansatz, and efforts have been made towards exploring different choices of ansatz such as unitary coupled cluster,\cite{VQE} hardware-efficient (heuristic) ansatze,\cite{Kandala2017} ADAPT-VQE,\cite{Grimsley2019} and qubit coupled cluster,\cite{Ryabinkin2018} among others. A step towards practical quantum algorithms is to use a hardware-efficient ansatz along with qubit operators to improve the energetics, resulting in a non-unitary VQE algorithm named LAS-nuVQE.\cite{wang2025nonunitaryvariationalquantumeigensolver} One limitation of the hardware-efficient ansatz is the lack of spin-conservation, which can be overcome through the use of a spin penalty. Measurement cost mitigation methods have also been explored in reference \citenum{wang2025nonunitaryvariationalquantumeigensolver}.

The algorithms discussed above can be improved; for example, by the careful consideration of spin preservation or additional bootstrapping of quantum methods using information from classical quantum chemistry. In addition, fragmentation and embedding methods continue to be used to extend the reach of quantum algorithms. In the future, quantum algorithms are poised to have a significant impact on cutting-edge applications such as the prediction of experimental spectra via modeling excited states \cite{gocho2023excited,higgott2019variational,cadi2024folded} and understanding the quantum dynamics of complex systems~\cite{roggero2020preparation,ollitrault2021molecular}. 

\section{Summary and outlook}
In summary, we discussed quantum embedding algorithms, for single- and multireference cases, utilizing classical and potentially also quantum computers. These methods have been employed to explore a broad spectrum of problems characterized by static correlation, such as excited states in spin defects, band structures, adsorption processes, bond dissociation, and spin-state energetics. 

One of the future directions is to expand the application of these methodologies to the field of heterogeneous catalysis \cite{hariharan2024modeling}, especially for systems that involve transition metal elements with strong electron correlation. However, there are instances where the methods developed so far are not sufficient. For example, semiconductor nanocrystals can undergo nonradiative recombination through defect-induced conical intersections that lowers energy conversion efficiency in optoelectronics~\cite{ShuDefect2015,ShuUnderstanding2017, ShuFirst2016}. While multireference DMET would be an optimal choice for modeling such potential energy surfaces, as excited states are inherently multireference, methods such as NEVPT2-DMET or DME-PDFT are single-state methods and are inaccurate in regions of strong nuclear-electron coupling, such as conical intersections or locally avoided crossings. 
Multi-state versions of NEVPT2 and MC-PDFT, such as quasi-degenerate NEVPT2 (QD-NEVPT2) \cite{Angeliquasidegenerate2004} or the recently developed linearized pair-density functional theory (L-PDFT), exist
\cite{HennefarthLinearized2023, HennefarthLinearized2023a}; however, their integration into the DMET framework has yet to be developed and tested. One would have to ensure that the final energies for each state come from the diagonalization of an effective Hermitian matrix to avoid non-physical state crossings in regions of strong nuclear-electron couplings. Hence, the development of multi-state multireference DMET would allow for accurate modeling of the potential energy surface of extended systems near conical intersections and locally avoided crossings, which are topological features that drive internal conversion and nonradiative decay processes. Recently, MC-PDFT using a RASSCF wave function has shown promising performance for the simulation of the X-ray absorption spectroscopy \cite{ghosh2023combined} and non-resonant Auger spectroscopy \cite{fouda2025computation}. These results suggest that embedded multireference methods could be effectively applied to core-level spectroscopy of molecules adsorbed on surfaces. Additionally, systems requiring large active spaces may be treated efficiently using localized active space approaches~\cite{LASSCF1, hermes2025localized}.
In the future it will also be interesting to use embedding methods to explore spin-phonon coupling in periodic systems, in analogy with what can be currently done for middle-size molecules~\cite{haldar2025role}.
Recent work combining quantum embedding methods and hybrid quantum-classical algorithms has shown promise in the sizes of the active spaces that can be treated. 
The advantages promised by future quantum hardware have been explored to develop quantum algorithms based on the fragment-based LASSCF method, such as LAS-UCCSD, its selected analogue LAS-USCCSD, LAS-QKSD, and LAS-nuVQE where the interfragment electron correlation is restored using quantum algorithms. 
A key challenge moving forward for both multireference and single-reference quantum algorithms is the development of robust, method-specific noise mitigation strategies. Encouragingly, with ongoing progress in this area, the implementation of such techniques appears feasible in the near future~\cite{endo_practical_2018,bravyi_mitigating_2021}.

\begin{acknowledgement}
This material is based upon work supported by the U.S. Department of Energy, Office of Science, National Quantum Information Science Research Centers. M.R. Hermes and V.A. are partially supported by the U.S. DOE, Office of Basic Energy Sciences, Division of Chemical Sciences, Geosciences, and Biosciences under Award Number USDOE/DE-SC002183 and by the U.S. DOE, Office of Basic Energy Sciences, Heavy Elements Chemistry Program, under Award Number DE-SC0023693. Q.W. and R.D. are partially supported by IBM-UChicago Quantum Collaboration, under agreement number MAS000364. B.J. and S.V. are partially supported by the AFOSR, under Award Number FA9550-20-1-0360. M.R. Hennefarth acknowledges support by the National Science Foundation Graduate Research Fellowship under Grant No. 2140001. Any opinions, findings, conclusions, or recommendations expressed in this material are those of the author(s) and do not necessarily reflect the views of the National Science Foundation.
\end{acknowledgement}

\bibliography{ref}

\end{document}